\documentstyle[amsmath,amssymb,graphicx]{article}

\def\be{\begin{eqnarray}}
\def\ee{\end{eqnarray}}
\def\nn{\nonumber}

\def\p{\partial}
\def\tr{{\rm tr}\,}

\def\la{\ \langle}
\def\ra{\rangle\ }
\def\lla{\ \langle\langle}
\def\rra{\rangle\rangle\ }
\def\bla{\ \Big<}
\def\bra{\Big>\ }
\def\blla{\ \Big<\Big<}
\def\brra{\Big>\Big>\ }

\def\K{C^{{\rm conn}}}
\def\rhoc{\rho^{{\rm conn}}}

\def\e{e^{{\rm conn}}}

\hoffset= - 1.0in         % switch off for draft style
\begin{document}

\hfill ITEP/TH-23/10

\bigskip

\centerline{\Large{From Brezin-Hikami to Harer-Zagier formulas for Gaussian correlators
}}

\bigskip

\centerline{\it A.Morozov and Sh.Shakirov}

\bigskip

\centerline{ITEP, Moscow, Russia}

\bigskip

\centerline{ABSTRACT}

\bigskip

{\footnotesize
Brezin-Hikami contour-integral representation
of exponential multidensities in finite $N$ Hermitian matrix model
is a remarkable implication of the old
Hermitian-Kontsevich duality.
It is also a simplified version of Okounkov's formulas
for the same multidensities in the cubic Kontsevich model
and of Nekrasov calculus for LMNS integrals,
a central piece of the modern studies of AGT relations.
In this paper we use Brezin-Hikami representation to
derive explicit expressions for the Harer-Zagier multidensities
(from arXiv:0906.0036):
the only known exhaustive generating functions
of all-genera Gaussian correlators which are fully calculable
and expressed in terms of elementary functions. Using the Brezin-Hikami contour integrals, we rederive the 1-point function of Harer and Zagier and the 2-point arctangent function of arXiv:0906.0036. We also present (without a proof) the explicit expression for the 3-point function in terms of arctangents. Derivation of the 3-point and higher Harer-Zagier functions remains a challenging problem. }

\bigskip

\tableofcontents

\section{Introduction}

Gaussian correlators are among the central quantities
in mathematical physics.
For many decades they remain the subject of constant investigation
and innumerable applications,
still we are far from a satisfactory and exhaustive description.
The problem sounds very simple: to evaluate
Gaussian integrals over $N\times N$ Hermitian matrices
\be
C_{i_1\ldots i_k}(N) = \la \tr \phi^{i_1} \ldots \tr \phi_{i_k}\ra
\equiv \frac{\int_{N\times N}  \tr \phi^{i_1} \ldots \tr \phi_{i_k}
e^{-\frac{1}{2}\tr \phi^2} d\phi}{
\int_{N\times N}  e^{-\frac{1}{2}\tr \phi^2} d\phi}
\ee
or their {\it connected} counterparts
\be
\K_{i} = \lla \tr \phi^{i} \rra = C_i
\ee

\be
\K_{ij} = \lla \tr \phi^{i} \tr \phi^{j} \rra = C_{ij} - C_iC_j
\ee

\be
\K_{ijk} = \lla \tr \phi^{i} \tr \phi^{j} \tr \phi^{k} \rra = C_{ijk} - C_{ij}C_k - C_{ik}C_j - C_{jk}C_i + 2 C_i C_j C_k
\ee

$$ \ldots \ldots \ldots $$
and it is an easy exercise to find any particular quantity of this kind, for particular low values of $i_1, \ldots, i_k$
\be
C_0(N) = N \nn\\
C_2(N) = N^2 \nn\\
C_4(N) = 2 N^3 + N, \nn\\
C_6(N) = 5 N^4 + 10 N^2, \nn\\
C_{22}(N) = N^4 + 2 N^2, \nn\\
C_{222}(N) = N^6 + 6 N^4 + 8 N^2, \nn\\
\ee
or, say, for particular low values of $N$:

\be
C_{k_1, \ldots, k_m}(1) = (k_1 + \ldots + k_m - 1)!! \ \ \mbox{ if } k_1 + \ldots + k_m \mbox{ is even, and 0 otherwise }
\ee
\smallskip\\
The fact that each connected correlator is a non-trivial polynomial in $N$
is encoded in the idea of a {\it genus expansion}: different powers of $N$
come from the fat-graph Feynman diagrams of different topology.
The real problem is to find a closed expression for {\it generic}
$C$ and $\K$ or, at least, for contributions of particular genera.

As usual, straightforward approach is to construct a generating function
and evaluate it exactly.
Also as usual, resolvability of this problem depends on the {\it clever}
choice of the generating function.
The simplest choice is to consider {\it resolvents}
\be
\rho(z_1,\ldots,z_k) = \frac{\nabla(z_1)\ldots\nabla(z_k) {\cal Z}}{{\cal Z}}
\equiv \sum_{i_1,\ldots,i_k=0}^\infty
\frac{C_{i_1\ldots i_k}(N)}{ z_1^{i_1+1}\ldots z_k^{i_k+1}} =
\bla \tr\frac{1}{z_1 - \phi}\ldots \tr\frac{1}{z_k-\phi}\bra
\ee
or
\be
\rhoc(z_1,\ldots,z_k) = \nabla(z_1)\ldots\nabla(z_k)\log{\cal Z}
\equiv \sum_{i_1,\ldots,i_k=0}^\infty
\frac{\K_{i_1\ldots i_k}(N)}{ z_1^{i_1+1}\ldots z_k^{i_k+1}} =
\blla \tr\frac{1}{z_1 - \phi}\ldots \tr\frac{1}{z_k-\phi}\brra
\ee
These generating functions are very well adjusted to the use of
Ward indentities for Gaussian matrix integrals, also known
under the names of loop equations or Virasoro constraints,
see \cite{UFN3} for a comprehensive review.
The problem is, however, that only contributions of particular
genera can be calculated in a closed form this way \cite{AMM1}:
\be
\rhoc_0(z) = \frac{z-y_N(z)}{2}, \ \ \
\rhoc_1(z) = \frac{N}{y_N^5(z)}, \ \ \
\rhoc_2(z) = \frac{21N(z^2+N)}{y_N^{11}(z)}, \ \ \ \nn \\
\rhoc_0(z_1,z_2) = \dfrac{1}{2(z_1-z_2)^2}\left(\dfrac{z_1z_2 - 4N}{y_N(z_1)y_N(z_2)} - 1 \right), \nn \\
\rhoc_0(z_1,z_2,z_3) = \dfrac{2N(z_1z_2+z_1z_3+z_2z_3+4N)}{y_N(z_1)^3 y_N(z_2)^3 y_N(z_3)^3}
\ee
where $y_N^2(z) = z^2-4N$ is what is called a spectral curve of the (in this case, Gaussian) matrix model. Given a spectral curve,
Virasoro constraints can be re-interpreted as
"topological recursion" rules \cite{2,EyOr,AMM},
allowing to build $\rho_g(z_1,\ldots,z_n)$ one after
another, recursively in the numbers of handles and
punctures $g$ and $n$.

As one calculates more and more resolvents of particular genera, it starts getting clear that formulas become increasingly complicated: it is rather hard to imagine that generic expression for $\rho$ or $\rhoc$ can be
obtained in any closed form. It turns out that a minor modification of the generating function
makes it calculable! Moreover, these {\it Harer-Zagier functions} turn
to be {\it elementary} functions.
In \cite{HZ} actually only a 1-point function was introduced
\be
\phi(z|N) \equiv \sum_{i=0}^\infty \frac{z^{2i}}{(2i-1)!!} \la \tr\phi^{2i}\ra
= \frac{1}{2z^2} \left(\left(\frac{1+z^2}{1-z^2}\right)^N - 1\right)
\ee
and the difference from $\rho(z)$ is the peculiar bifactorial in the denominator.
As shown in \cite{MS_HZ}, this result is by no means limited to
a 1-point function: all Harer-Zagier functions $\phi(z_1,\ldots,z_n)$
look comprehensible!
To make expressions really elementary one should also get rid of $N$,
by making additional generating function \cite{AMM1, MS_HZ}:
\be
\hat\phi(z|\lambda) \equiv \sum_{N=0}^\infty \phi(z|N) \lambda^N =
\frac{1}{2z^2}\left( \frac{1}{1 - \lambda\frac{1+z^2}{1-z^2}}
- \frac{1}{1-\lambda}\right) =
\frac{\lambda}{(1-\lambda)}\frac{1}{1-\lambda - (1+\lambda)z^2}
\label{Phi1p}
\ee
Then, for example \cite{MS_HZ},
\be
\hat\phi_{\rm odd}(z_1,z_2|\lambda) \equiv
\sum_{N=0}^\infty \lambda^N \sum_{i_1,i_2=0}^\infty
\frac{z^{2i_1+1}}{(2i_1+1)!!}\frac{z^{2i_2+1}}{(2i_2+1)!!}\ C_{2i_1+1,2i_2+1}(N)
= \frac{\lambda}{(1-\lambda)^{3/2}}
\frac{\arctan\frac{z_1z_2\sqrt{1-\lambda}}{\sqrt{1-\lambda + (1+\lambda)(z_1^2+z_2^2)}}}
{\sqrt{1-\lambda + (1+\lambda)(z_1^2+z_2^2)}}
\label{Phi2podd}
\ee
and so on, see s.\ref{HZR} for more examples. The purpose of this paper is to shed new light to calculability of such Harer-Zagier functions and make a step towards derivation of closed and generic
expressions for {\it all} of them. The idea is to use a still another, third, type of the generating function,
\be
e(s_1,\ldots,s_k) = \bla \tr e^{s_1\phi}\ldots \tr e^{s_k\phi}\bra
\ee
and
\be
\e(s_1,\ldots,s_k) = \blla \tr e^{s_1\phi}\ldots \tr e^{s_k\phi}\brra
\ee
They turn out to be closely related to (Generalized) {\it Kontsevich matrix model}
(GKM) \cite{GKM,UFN3},
which in Gaussian case possesses a remarkable {\it duality} \cite{dua} with
the Gaussian matrix model, considered in the present paper.
This duality led E.Breain and S.Hikami to a wonderful representation
of $e(s|N)$ in the contour integral form \cite{BH}:
\be
e(s_1,\ldots,s_k|N) =
\prod_{i=1}^k \frac{e^{\frac{s_i^2}{2}}}{s_i} \oint du_i e^{u_is_i}
\left(1+\frac{s_i}{u_i}\right)^N
\det_{1\leq i,j\leq k} \frac{1}{u_i - (u_j+s_j)}
\label{BHfla}
\ee
The last determinantal factor here can be also rewritten as
\be
\det_{1\leq i,j\leq k} \frac{1}{u_i - (u_j+s_j)} = %(-)^{N(N-1)/2}???
\prod_{i>j} \frac{(u_i-u_j)(u_i-u_j+s_i-s_j)}{(u_i-u_j+s_i)(u_i-u_j-s_j)}
\ee
what makes the whole expression similar to LMNS integral \cite{LMNS},
which recently attracted a lot of new attention in the context of
AGT-relations \cite{AGT}: as already emphasized in \cite{AGTdua}
duality of \cite{dua,BH} should play a prominent role in the understanding
of AGT relations, at least on the lines of the Dotsenko-Fateev-matrix-model
approach \cite{DFMM}.
Remarkably, eq.(\ref{BHfla}) possesses a direct generalization at least
to {\it cubic} Kontsevich model \cite{Ok}, and questions now arise
about further generalizations to arbitrary GKM and about the possible AGT-like
applications of these generalizations.

In this paper, however, we do not proceed in these very appealing directions.
Instead, we concentrate on the application of (\ref{BHfla}) to
evaluation of Harer-Zagier functions, which
-- in variance with $\rho(z)$ and $e(s)$ --
are elementary fully calculable functions.
Application of (\ref{BHfla}) for this purpose will be absolutely straightforward.
First, conversion from $N$ to $\lambda$ substitutes the factor

$$\prod_{i=1}^k \left(1+\frac{s_i}{u_i}\right)^N$$
by

$$\left(1 - \lambda  \prod_{i=1}^k \left(1+\frac{s_i}{u_i}\right)\right)^{-1}$$
Second, contour integrals just pick up contributions from poles
of the simple rational function of $u_i$ -- in a very similar way to
Nekrasov's celebrated calculation \cite{Nek}.
Third, the Harer-Zagier bifactorials are made from factorials in $e^{s\phi}$
by shifting $s_i\rightarrow s_it_i$ and introducing additional
semi-integer-$\Gamma$-function factors with the help of $t$-integrations
$\int_0^\infty e^{-t_i^2/2}dt_i$.

In this way we reproduce the 1-point and 2-point Harer-Zagier functions from the Brezin-Hikami integrals. Generalization to 3-point and higher functions should be straightforward, but a clever method needs to be invented to convert the resulting multiple integrals into elementary functions (which, as we conjectured in \cite{MS_HZ}, are always elementary functions of arctangent type). This conversion is trivial in the 1-point case; in the 2-point case for the purpose of such conversion we use series representations, which generalize the elementary formula

\be
\frac{\arctan\frac{c}{\sqrt{ab-c^2}}}{\sqrt{ab-c^2}} = \sum_k \frac{4^k (k!)^2}{(2k+1)!}\frac{c^{2k+1}}{(ab)^{k+1}} = \int_0^\infty \int_0^\infty  e^{-at_1^2-bt_2^2}\sin(2ct_1t_2) dt_1dt_2
\ee
see s.3.2. It is not yet clear if this series approach is really
promissing for 3-point and higher cases. Probably, some new ideas are required to reveal the general structure of $k$-point Harer-Zagier elementary functions for $k \geq 3$.

\pagebreak

\section{Harer-Zagier formulas: a reminder
\label{HZR}}

Harer-Zagier functions (multidensities) in the finite $N$ Gaussian model are defined as

\be
\phi(s_1, \ldots, s_k|N) = \sum_{i_1,\ldots,i_k=0}^\infty \K_{i_1\ldots i_k}(N) \dfrac{s_1^{i_1}}{n(i_1)} \ldots \dfrac{s_k^{i_k}}{n(i_k)}
\ee
(we will consider only connected Harer-Zagier functions here) where the bifactorial normalisation is

\be
n(i) = \left\{ \begin{array}{cc} (i-1)!!, \ \ \ i = \mbox{ even } \\ i!!, \ \ \ i = \mbox{ odd } \\ \end{array} \right.
\ee
As emphasized in the previous section, it turns out to be convenient to take a generating function with respect to the size of a matrix $N$ as well, converting it into a continuous parameter $\lambda$:

\be
\hat\phi(s_1, \ldots, s_k|\lambda) = \sum\limits_{N = 0}^{\infty} \lambda^N \phi(s_1, \ldots, s_k|N)
\ee
Technically, this conversion is done with the following well-known finite and infinite summation formulas: since
\be
\frac{1}{(1-\lambda)^{k+1}} =
\sum_N \frac{(N+k)!}{N!k!}\lambda^N
%\sum_N \frac{N(N+1)\ldots (N+k)}{(k+1)!}\lambda^N,
\label{invdegexp}
\ee
we have:
\be
\sum_N \lambda^N = \frac{1}{1-\lambda}, \nn \\
\sum_N N\lambda^N = \frac{\lambda}{(1-\lambda)^2}, \nn \\
\sum_N N^2\lambda^N = \frac{\lambda(1+\lambda)}{(1-\lambda)^3}, \nn \\
\sum_N N^3\lambda^N = \frac{\lambda(1+4\lambda+\lambda^2)}{(1-\lambda)^4}, \nn \\
\sum_N N^4\lambda^N = \frac{\lambda(1+11\lambda+11\lambda^2+\lambda^3)}{(1-\lambda)^5}, \nn \\
\ldots
\label{lambdasers}
\ee
Similarly, from an obvious formula
\be
\sum_{j=0}^n j(j-1)(j-2)\ldots (j-m) =
\frac{(n+1)n(n-1)(n-2)\ldots (n-m)}{m+2}
\ee
(l.h.s. is a polynomial of degree $m+1$ in $n$,
which obviously vanishes at $n=0,1,2,\ldots,m$)
it follows that
\be
\sum_{j=0}^n j = \frac{n(n+1)}{2},\nn \\
\sum_{j=0}^n j^2 = \frac{n(n+1)(2n+1)}{6},\nn \\
\sum_{j=0}^n j^3 = \frac{n^2(n+1)^2}{4},\nn \\
\sum_{j=0}^n j^4 = \frac{n(n+1)(2n+1)(3n^2+3n-1)}{30}
\ldots
\ee

\subsection{1-point function}

The 1-point Harer-Zagier density has a form \cite{AMM1, HZ, MS_HZ}

\be
\hat\phi(s|\lambda) = \frac{\lambda}{(1-\lambda)}\frac{1}{1-\lambda - (1+\lambda)s^2}
\ee
This simple and beautiful formula admits generalizations to 2-point and higher functions, see \cite{MS_HZ} and below.

\subsection{2-point function}

The 2-point Harer-Zagier density has a form

\be
\hat\phi(s_1, s_2|\lambda) = \hat\phi_{even}(s_1, s_2|\lambda) + \hat\phi_{odd}(s_1, s_2|\lambda)
\ee
where a convenient separation of the even and odd parts is introduced, and

\be
\hat\phi_{odd}(s_1, s_2|\lambda) = \frac{\lambda}{(1-\lambda)^{3/2}}
\frac{\arctan\frac{z_1z_2\sqrt{1-\lambda}}{\sqrt{1-\lambda + (1+\lambda)(z_1^2+z_2^2)}}}
{\sqrt{1-\lambda + (1+\lambda)(z_1^2+z_2^2)}}
\ee
\be
\hat\phi_{even}(s_1, s_2|\lambda) = \dfrac{s_1s_2}{s_1^2-s_2^2}\left(s_1\dfrac{\partial}{\partial s_1} - s_2\dfrac{\partial}{\partial s_2}\right)\hat\phi_{odd}(s_1, s_2|\lambda)
\ee
These formulas were derived in \cite{MS_HZ} with the help of integrable Toda equations. In the present paper we rederive them, using a different method (Brezin-Hikami contour integrals) see below.

\subsection{3-point function}

The 3-point Harer-Zagier density has a form

\be
\hat\phi(s_1, s_2, s_3|\lambda) = \hat\phi_{even}(s_1, s_2, s_3|\lambda) + \hat\phi_{mix}(s_1, s_2, s_3|\lambda)
\ee
where $\hat\phi_{even}(s_1, s_2, s_3|\lambda)$ is a totally even (even w.r.t each of its arguments) part of the function. We present the following expression for this totally even part (without a proof, at conjectural level):

\begin{align}
\nonumber f\big(\tau \mid s_1, s_2, s_3 \big) \ = \ & \left(
\dfrac{1 - s_1}{(1 + \tau^2 s_1 (s_2 + s_3))^2} + \dfrac{1 + s_2 + s_3 - s_1}{1 + \tau^2 s_1 (s_2 + s_3)}
\right) \dfrac{{\rm arctan}\left( \sqrt{\dfrac{s_2 s_3(1 - s_1) \tau^2}{1 + \tau^2 s_1 (s_2 + s_3)}} \right)}{\sqrt{\dfrac{s_2 s_3(1 - s_1) \tau^2}{1 + \tau^2 s_1 (s_2 + s_3)}}} + \emph{} \\ \nonumber & \\ \nonumber & \emph{} +
\left(
\dfrac{1 - s_2}{(1 + \tau^2 s_2 (s_1 + s_3))^2} + \dfrac{1 + s_1 + s_3 - s_2}{1 + \tau^2 s_2 (s_1 + s_3)}
\right) \dfrac{{\rm arctan}\left( \sqrt{\dfrac{s_1 s_3(1 - s_2) \tau^2}{1 + \tau^2 s_2 (s_1 + s_3)}} \right)}{\sqrt{\dfrac{s_1 s_3(1 - s_2) \tau^2}{1 + \tau^2 s_2 (s_1 + s_3)}}} + \emph{} \\ \nonumber & \\ \nonumber &
\emph{} + \left(
\dfrac{1 - s_3}{(1 + \tau^2 s_3 (s_1 + s_2))^2} + \dfrac{1 + s_1 + s_2 - s_3}{1 + \tau^2 s_3 (s_1 + s_2)}
\right) \dfrac{{\rm arctan}\left( \sqrt{\dfrac{s_1 s_2(1 - s_3) \tau^2}{1 + \tau^2 s_3 (s_1 + s_2)}} \right)}{\sqrt{\dfrac{s_1 s_2(1 - s_3) \tau^2}{1 + \tau^2 s_3 (s_1 + s_2)}}} - \emph{} \\ \nonumber & \\ \nonumber & \emph{}
\end{align}
$$ - \dfrac{s_1(s_2 + s_3)}{s_2 s_3(1 + \tau^2 s_1 (s_2 + s_3))} - \dfrac{s_2(s_1 + s_3)}{s_1 s_3(1 + \tau^2 s_2 (s_1 + s_3))} - \dfrac{s_3(s_1 + s_2)}{s_1 s_2(1 + \tau^2 s_3 (s_1 + s_2))} + $$
\begin{align}
+\dfrac{(s_1 + s_2)(s_1 + s_3)(s_2 + s_3)}{s_1s_2s_3(1 + \tau^2 (s_1 s_2 + s_1 s_3 + s_2 s_3 - s_1 s_2 s_3)) }
\end{align}
with a notation

\be
f\big(\tau \mid s_1, s_2, s_3 \big) = \dfrac{4}{1 - \tau^2} \dfrac{(1-s_1-s_2-s_3)^3}{s_1s_2s_3} \ \hat\phi_{even}\left(Ts_1, Ts_2, Ts_3\Big|\dfrac{1-T}{1+T}\right), \ \ \ T = \tau \sqrt{1 - s_1 - s_2 - s_3}
\ee
This formula was not presented in \cite{MS_HZ}, but it can
also be deduced by the method of computer experiment,
exploited in that paper,
which becomes today a powerful approach to derivation of
important results and conjectures in mathematical physics. It would be interesting to prove (derive) this formula, by using either Toda equations
or Brezin-Hikami integrals.
It is even more interesting to generalize to further generalize
it to the 4-point and higher functions, and check the appealing
conjecture that {\it all} the Harer-Zagier functions are
expressed through actangents, as linear combinations,
reminiscent of the Okounkov's eq. (\ref{Okoun}), which we remind in
the Appendix.

\section{Harer-Zagier formulas from Brezin-Hikami integrals}

\subsection{1-point function}

In the case of a 1-point function (\ref{BHfla}) is especially
easy to derive \cite{BH},
however already in this case the calculation
is actually done in Gaussian Kontsevich model, not in
Hermitian model itself -- thus emphasizing the role
of Hermitian-Kontsevich duality \cite{dua}.
\be
\frac{1}{V_N}\int_{N\times N} \tr e^{s\phi} e^{-\tr\phi^2/2} e^{\tr A\phi} d\phi
= \sum_k \int e^{s\phi_k} \Delta^2(\phi) \frac{\det e^{\phi_iA_j}}{\Delta(\phi)\Delta(A)}
\prod_i e^{-\phi_i^2/2}d\phi_i = \nn \\
= \frac{1}{\Delta(A)} \sum_k \int \Delta(\phi) e^{s\phi_k}
\sum_P (-)^P \prod_i e^{-\phi_i^2/2 + \phi_i A_{P(i)}} d\phi_i
= \frac{1}{V_N}\sum_k \frac{\Delta(\tilde A_{(k)})}{\Delta(A)} e^{\tr\tilde A_{(k)}^2}
\label{BHgkm}
\ee
where $V_N = {\rm Vol}_{SU(N)}$ is the volume of unitary group
and $\tilde A$ is the matrix with eigenvalues
\be
\tilde A_i^{(k)} = A_i + s\delta_{ki}
\ee
This is an elementary essentially rational expression,
however, it looks singular in the limit of $A\rightarrow 0$.
The limit can be easily taken if the formula is rewritten in
a somewhat more complicated form of a contour integral \cite{BH}:
\be
\sum_k \prod_{i<j}\frac{a_i-a_j + s(\delta_{ik}-\delta_{jk})}{a_i-a_j}
\prod_i e^{(a_i + s\delta_{ik})^2/2} =
e^{s^2/2}\prod_i e^{a_i^2/2} \sum_k e^{sa_k} \prod_{i\neq k}
\frac{a_i-a_k-s}{a_i-a_k} = \nn \\
= e^{s^2/2} \prod_i e^{a_i^2/2}
\oint e^{us} du \prod_i \frac{u - a_i + s}{u-a_i}
\ee
For $A=0$ this turns into
\be
\frac{1}{s}e^{s^2/2} \oint e^{us} \left(1+\frac{s}{u}\right)^N du
\label{1pBH}
\ee

Following \cite{MS_HZ} one can now pass to a generating function
in $N$: $f(N)\rightarrow \hat f(\lambda) = \sum_N \lambda^N f(N)$.
This could be done in different ways, for example, one could
write $\sum_N \frac{\lambda^N}{N!}f(N)$ instead.
Our choice, however, makes calculation of contour integrals
most convenient and elementary. Indeed, instead of (\ref{1pBH})
we obtain:
\be
%\boxed{
\hat e(s|\lambda) = \frac{e^{s^2/2}}{s}
\oint \frac{e^{us}du}{1 - \lambda\left(1+\frac{s}{u}\right)}
= \boxed{
\frac{\lambda}{(1-\lambda)^2}
e^{\frac{1+\lambda}{1-\lambda}\frac{s^2}{2}}
}
\label{E1p}
\ee
where contour integral picks up a contribution from a single
pole at $u = \frac{\lambda s}{1-\lambda}$.

Harer-Zagier function differs by the substitution
\be
e(s) = \sum_k \frac{s^{2k}}{(2k)!}\left<\tr \phi^{2k}\right>
\ \longrightarrow\
\phi(s) = \sum_k \frac{s^{2k}}{(2k-1)!!}\left<\tr \phi^{2k}\right>
\ee
Since
\be
\frac{1}{(2k-1)!!} = \frac{2^k k!}{(2k)!} =
\frac{1}{(2k)!}\int_0^\infty t^{2k+1} e^{-t^2/2} dt
\ee
we have
\be
%\boxed{
\hat\phi(s|\lambda) = \int_0^\infty \hat e(st|\lambda) e^{-t^2/2} tdt
= \frac{\lambda}{(1-\lambda)^2} \int_0^\infty
e^{-\frac{1}{2}t^2\left(1-\frac{1+\lambda}{1-\lambda}s^2\right)} tdt =
\boxed{
\frac{\lambda}{(1-\lambda)\Big(1-\lambda - (1+\lambda)s^2\Big)}
}
\ee
what successfully reproduces (\ref{Phi1p}).

Note that in this case both exponential and Harer-Zagier
generating functions are elementary.
In fact from (\ref{E1p}) one can also obtain an explicit
formula for the 1-point Gaussian resolvent:

\be
\hat\rho(z|\lambda) \equiv \left<\tr\frac{1}{z-\phi}\right>
= \int_0^\infty \hat e(s|\lambda) e^{-sz} ds =
\frac{\lambda}{(1-\lambda)^2}\int_0^\infty
e^{\frac{1+\lambda}{1-\lambda}\frac{s^2}{2} - sz}ds
= \frac{i\lambda}{(1-\lambda)\sqrt{1-\lambda^2}}\
{\rm erf}\left(iz\sqrt{\frac{1-\lambda}{1+\lambda}}\right)
\ee
\smallskip\\
We need here an expansion of error-function

\be
{\rm erf}(z) \equiv \int_0^\infty e^{-\frac{x^2}{2}-xz} dx
= e^{\frac{z^2}{2}}\int_z^\infty e^{-\frac{x^2}{2}} dx
\ \stackrel{x^2=z^2+t}{=}\
\frac{1}{z}\int_0^\infty \frac{e^{-t}dt}{\sqrt{1+\frac{2t}{z^2}}}
\ee
\smallskip\\
at {\it large} values of its argument, where it becomes
a series in inverse powers of $z$.
Adequate for working out this asymptotics is
the last representation, which immediately implies:

\be
\boxed{
\hat\rho(z|\lambda) =
\sum_{k=0} \frac{\lambda(1+\lambda)^k}{(1-\lambda)^{k+2}}
\frac{(2k-1)!!}{z^{2k+1}}
}
\label{Rho1p}
\ee
\smallskip\\
Using (\ref{lambdasers}) one can easily check the
consistency of this formula with the well known
first terms of the $\rho$-expansion:

\be
\rho(z) = \frac{<\tr I>}{z} + \frac{<\tr \phi^2>}{z^3}
+ \frac{<\tr\phi^4>}{z^5} + \ldots
= \frac{N}{z} + \frac{N^2}{z^3} + \frac{2N^3+N}{z^5} + \ldots
\ee
\smallskip\\
It is also clear from (\ref{Rho1p}) that the resolvent,
including the contributions of all genera is represented
by asymptotic, nowhere convergent series, and is not expressed through elementary functions.
Expressible are particular terms $\rho^{(p|1)}(z)$
of genus expansion \cite{AMM1},

\be
\rho(z) = \frac{z-y(z)}{2} + \frac{N}{y^5(z)}
+ \frac{21N(z^2+N)}{y^{11}(z)}
+ \frac{11N(135z^4+558Nz^2+158N^2)}{y^{17}(z)} + \ldots
= \frac{N}{z} + \frac{N^2}{z^3} + \frac{2N^3+N}{z^5} + \ldots
\label{rhozstan}
\ee
\smallskip\\
with $y^2(z) = z^2-4N$.
Particular terms of the genus expansion (\ref{rhozstan}) can be
directly deduced from the general formula (\ref{Rho1p}).
For this it should be re-expanded in monomials $(1-\lambda)^{-s}$:
\be
\hat\rho(z|\lambda) =
\sum_{k=0} \frac{\lambda(1+\lambda)^k}{(1-\lambda)^{k+2}}
\frac{(2k-1)!!}{z^{2k+1}} =
\sum_k \sum_{s=0}^\infty
\frac{(2k-1)!!}{z^{2k+1}} \frac{2^k}{(1-\lambda)^{s+1}}
\oint_0 \frac{(1-\xi)(1-\xi/2)^k}{\xi^{k+2}}\xi^sd\xi = \nn \\
\stackrel{(\ref{invdegexp})}{=}\
\sum_k \frac{2^k(2k-1)!!}{z^{2k+1}} \sum_s \sum_N
\frac{(N+s)!}{N!s!} \lambda^N
\oint_0 \frac{(1-\xi)(1-\xi/2)^k}{\xi^{k+2}}\xi^sd\xi \
\stackrel{(\ref{invdegexp})}{=}\
\sum_k \frac{2^k(2k-1)!!}{z^{2k+1}} \sum_N \lambda^N
\oint_0 \frac{(1-\xi/2)^k}{(1-\xi)^N\xi^{k+2}}d\xi =
\nn \\
= \sum_N \lambda^N \sum_k \frac{2^k(2k-1)!!}{z^{2k+1}(k+1)!}
\left.\p_\xi^{k+1}\frac{(1-\xi/2)^k}{(1-\xi)^N}\right|_{\xi =0} = \nn \\
= \sum_{N}\lambda^N \sum_k \frac{2^k(2k-1)!!}{z^{2k+1}}
\left(\frac{N^{k+1}}{(k+1)!} + \frac{N^{k-1}}{12(k-2)!} +
\frac{N^{k-3}(5k-2)}{1440(k-4)!} +
\frac{N^{k-5}(35k^2-77k+12)}{128\cdot 81\cdot 35\cdot (k-6)!} +
\ldots \right) = \nn \\
= \sum_{N}\lambda^N \sum_k \frac{2^k(2k-1)!!}{z^{2k+1}}
\left\{\frac{N^{k+1}}{(k+1)!} + \frac{N^{k-1}}{12(k-2)!} +
\left(\frac{N^{k-3}}{288(k-5)!} + \frac{N^{k-3}}{80(k-4)!}\right)
+ \right. \nn \\ \left.
+ \left(\frac{N^{k-5}}{128\cdot 81\cdot (k-8)!}
+ \frac{N^{k-5}}{960\cdot (k-7)!}
+ \frac{N^{k-5}}{64\cdot 7\cdot (k-6)!}\right)
+ \ldots \right\} = \nn \\
= \sum_N \lambda^N \left(\frac{z-y(z)}{2} + \frac{N}{y^5(z)}
+ \frac{21N(z^2+N)}{y^{11}(z)}
+ \frac{11N(135z^4+558Nz^2+158N^2)}{y^{17}(z)} + \ldots\right)
\ee
In this manipulation eq.(\ref{invdegexp}) is used twice:
first in order to convert $(1-\lambda)^{-s-1}$ into a power series
in $\lambda$ and second to find a sum over $s$, which is equal
to $(1-\xi)^{-N-1}$.
The integration variable $\xi$ is similar to $1-\lambda$.

Starting from 2-point functions, only Harer-Zagier
functions are expressible in terms of elementary functions.
Like more general Okounkov's formulas for cubic Kontsevich
model, expressions for exponential correlators remain
well-defined (convergent), but can be reduced only to
integrals of elementary functions. Transition from resolvents to convergent exponential correlators
is nothing but the application of the standard Pade summation method,
while transition to Harer-Zagier correlators is a fancy modification
of Pade method, leading -- at least in the case of Gaussian correlators --
not only to {\it convergent}, but to {\it elementary} expressions. In fact transition between the Harer-Zagier and exponential functions
is also an ordinary Pade transform, only with a twice smaller parameter:
\be
\hat\rho(z) = \frac{A}{z} \sum_k (2k-1)!!B^k\ \ \longrightarrow\ \
\hat\phi(s) = A \sum_k B^k = \frac{A}{1-B}\ \  \longrightarrow\ \
\hat e(s) = A \sum_k \frac{B^k}{2^kk!} = Ae^{B/2}
\ee
with $A= \frac{\lambda}{(1-\lambda^2)}$,
$B = \frac{1+\lambda}{(1-\lambda)}s^2$
and $z = s^{-1}$.

\subsection{2-point function}

Expression (\ref{BHgkm}) remains almost the same for arbitrary
exponential correlators, only for a correlator of $n$
exponentials the label $k$ at the r.h.s. is
a vector of the size $n$ and
\be
e(s_1,\ldots,s_n|N) =
\frac{1}{V_N}\int_{N\times N} \prod_{\alpha=1}^n\tr e^{s_\alpha\phi}
e^{-\tr\phi^2/2} e^{\tr A\phi} d\phi
%= \sum_k \int e^{s\phi_k} \Delta^2(\phi) \frac{\det e^{\phi_iA_j}}{\Delta(\phi)\Delta(A)}
%\prod_i e^{-\phi_i^2/2}d\phi_i = \nn \\
%= \frac{1}{\Delta(A)} \sum_k \int \Delta(\phi) e^{s\phi_k}
%\sum_P (-)^P \prod_i e^{-\phi_i^2/2 + \phi_i A_{P(i)}} d\phi_i
= \frac{1}{V_N}\sum_{\vec k} \frac{\Delta(\tilde A_{(\vec k)})}
{\Delta(A)} e^{\tr\tilde A_{(\vec k)}^2}
\label{BHgkm}
\ee
and
\be
\tilde A_i^{(\vec k)} = A_i + \sum_{\alpha=1}^n s_\alpha\delta_{i k_\alpha }
\ee
Contour-integral representation, however, becomes a little more involved
\cite{BH}:
\be
e(s_1,\ldots,s_n|N) =
\prod_{\alpha=1}^n \frac{e^{s_\alpha^2/2}}{s_\alpha}\oint du_\alpha
e^{s_\alpha u_\alpha} \left(1+\frac{s_\alpha}{u_\alpha}\right)^N
\prod_{\alpha<\beta} \frac{(u_\alpha-u_\beta)(u_\alpha-u_\beta+s_\alpha-s_\beta)}
{(u_\alpha - u_\beta+s_\alpha)(u_\alpha-u_\beta-s_\beta)},
\ee
so that
\be
\hat e(s_1,\ldots,s_n|\lambda) =
\prod_{\alpha=1}^n \frac{e^{s_\alpha^2/2}}{s_\alpha}\oint du_\alpha
e^{s_\alpha u_\alpha}
\frac{\prod_\alpha u_\alpha}
{\prod_\alpha u_\alpha - \lambda\prod_\alpha (u_\alpha - s_\alpha)}
\prod_{\alpha<\beta} \frac{(u_\alpha-u_\beta)(u_\alpha-u_\beta+s_\alpha-s_\beta)}
{(u_\alpha - u_\beta+s_\alpha)(u_\alpha-u_\beta-s_\beta)}
\ee

In the case of a two-point function, $n=2$ the integral is two-fold:
\be
\hat e(s_1,s_2|\lambda) =
\frac{e^{(s_1^2+s_2^2)/2}}{s_1s_2}\oint\oint \frac{e^{s_1u+s_2v}  uv  du dv}
{\Big((1-\lambda)uv - \lambda us_2 - \lambda vs_1 - \lambda s_1s_2\Big)}
\frac{(u-v)(u-v+s_1-s_2)}{(u-v+s_1)(u-v-s_2)}
\ee
and the main difficulty comes from quadratic factor in the denominator.
It is actually linear both in $u$ and $v$, thus the first contour
integral, say, over $v$ is simple:
it picks up contributions from the three poles
$v = u + s_1$, $v=u-s_1$ and $v = \frac{\lambda s_2(u+s_1)}
{(1-\lambda)u - \lambda s_1}$.
However, the last of the three gives rise to a new pole in $u$,
and actually this is the only pole which contributes to the second
integral over $u$: contributions of other poles in $u$ cancel to zero.
After all these intermediate calculations, the resulting expression for the exponential 2-point function is
$$
\hat e(s_1,s_2|\lambda) =  \frac{s_1 \lambda^2}{(1-\lambda)^4}
\oint_{u = 0}\frac{du}{u^2}
\exp\left(\frac{(1+\lambda)(s_1^2+s_2^2)}{2(1-\lambda)} + {\frac{\lambda s_1s_2^2}{u(1-\lambda)^2}} + s_1u\right) \times
$$
$$
\times \frac{\left(1-\frac{u(1-\lambda)(s_1-s_2)}{s_1s_2}
-\frac{u^2(1-\lambda)^2}{\lambda s_1s_2}\right)
\left(1-\frac{u(1-\lambda)(s_1-s_2)}{\lambda s_1s_2}
-\frac{u^2(1-\lambda)^2}{\lambda s_1s_2}\right)}
{\left(1-\frac{u(1-\lambda)}{s_2}\right)
\left(1-\frac{u(1-\lambda)}{\lambda s_2}\right)}
$$
and it is not so easy to evaluate,
because $u^{-1}$ is present in the exponent. The problem is slightly simplified by taking the odd part of the function, $\hat e_{odd}(s_1,s_2|\lambda)$, which takes a somewhat more concise form
$$
\hat e_{odd}(s_1,s_2|\lambda) = \dfrac{1}{4} \Big( \hat e(s_1,s_2|\lambda) - \hat e(s_1,-s_2|\lambda) - \hat e(-s_1,s_2|\lambda) + \hat e(- s_1,-s_2|\lambda)\Big) =
$$

$$
= e^{\frac{(1+\lambda)(s_1^2+s_2^2)}{2(1-\lambda)}} \oint_{u = 0} du \dfrac{\lambda s_2 \big( u^2 (1 - \lambda)^2 + \lambda s_2^2 \big) }{ (\lambda^2 s_2^2 - u^2 (1 - \lambda)^2 )(s_2^2 - u^2 (1 - \lambda)^2 )} \sinh\left({\frac{\lambda s_1s_2^2}{u(1-\lambda)^2}} + s_1u\right)
$$
Rescaling then $u \mapsto u/s_1$ and denoting $S = s_1 s_2 / (1 - \lambda)$, one finds

$$
\hat e_{odd}(s_1,s_2|\lambda) = \dfrac{\lambda}{1-\lambda} e^{\frac{(1+\lambda)(s_1^2+s_2^2)}{2(1-\lambda)}} \oint_{u = 0} du \dfrac{S \big( u^2 + \lambda S^2 \big) }{ (\lambda^2 S^2 - u^2 )(S^2 - u^2 )} \sinh\left( \frac{\lambda S^2}{u} + u\right)
$$
Using the simple identity

$$\dfrac{S \big( u^2 + \lambda S^2 \big) }{ (\lambda^2 S^2 - u^2 )(S^2 - u^2 )} = \sum\limits_{k = 0}^{\infty} \dfrac{S^{2k + 1}}{u^{2k+2}} \dfrac{\lambda^{2k+1}-1}{\lambda-1}$$
we obtain a series

$$\hat e_{odd}(s_1,s_2|\lambda) = \dfrac{\lambda}{1-\lambda} e^{\frac{(1+\lambda)(s_1^2+s_2^2)}{2(1-\lambda)}} \sum\limits_{k = 0}^{\infty} \sum\limits_{m = 0}^{\infty} \dfrac{\lambda^{2k+1}-1}{\lambda-1} \dfrac{S^{2k + 2m + 1} \lambda^m}{m!(m+2k+1)!} $$
and this series can be used directly to calculate the Harer-Zagier function:

$$\hat\phi_{odd}(s_1,s_2) = \int\limits_{0}^{\infty} dt_1 \int\limits_{0}^{\infty} dt_2 e^{-t_1^2/2} e^{-t_2^2/2} \hat e_{odd}(s_1 t_1,s_2 t_2|\lambda) = $$

%$$\int\limits_{0}^{\infty} dt_1 \int\limits_{0}^{\infty} dt_2 e^{-t_1^2} e^{-t_2^2} \dfrac{\lambda}{1-\lambda} e^{\frac{(1+\lambda)(t_1^2s_1^2+t_2^2s_2^2)}{2(1-\lambda)}} \sum\limits_{k = 0}^{\infty} \sum\limits_{m = 0}^{\infty} \dfrac{\lambda^{2k+1}-1}{\lambda-1} \left( \dfrac{t_1 t_2 s_1 s_2}{1-\lambda} \right)^{2k + 2m + 1} \dfrac{(-1)^{m+k} \lambda^m}{m!(m+2k+1)!} = $$

$$= \dfrac{\lambda}{1-\lambda} \sum\limits_{k = 0}^{\infty} \sum\limits_{m = 0}^{\infty} \dfrac{\lambda^{2k+1}-1}{\lambda-1} \lambda^m 4^{k + m} \left( \dfrac{s_1 s_2}{1-\lambda} \right)^{2k + 2m + 1} \dfrac{(m+k)!(m+k)!}{m!(m+2k+1)!} \left( \dfrac{1}{1 - \frac{1+\lambda}{1-\lambda} s_1^2 } \cdot \dfrac{1}{1 - \frac{1+\lambda}{1-\lambda} s_2^2 } \right)^{k+m+1} = $$

\be
\boxed{
= \frac{\lambda}{(1-\lambda)^2}
\frac{\arctan\frac{s_1s_2}{\sqrt{1- \frac{1+\lambda}{1-\lambda}(s_1^2+s_2^2)}}}
{\sqrt{1- \frac{1+\lambda}{1-\lambda}(s_1^2+s_2^2)}}
}
\ee
In fact, the equality between the above double series and the 2-point arctangent function is not quite obvious and requires additional comments. A more natural series representation for the 2-point arctangent function is
\be
\frac{\arctan\frac{\gamma c}{\sqrt{ab-c^2}}}{\sqrt{ab-c^2}}
= \sum_k \frac{(\gamma c)^{2k+1}}{(2k+1)(ab-c^2)^{k+1}}
= \sum_{k,m} \frac{c^{2k2+2m+1}}{(ab)^{k+m+1}} \frac{(m+k)!}{m!k!}
\frac{\gamma^{2k+1}}{2k+1}
\ee
with compact notations $\gamma = (1-\lambda)/(1+\lambda)$, $a = 1 - \gamma s_1^2, b = 1 - \gamma s_2^2, c = s_1 s_2$. Comparing this natural double series with the above double series expansion of the contour integral, one observes a puzzling series identity

\be
4^p p! \sum_{k=0}^p \frac{(1-\lambda^{2k+1})\lambda^{p-k}}
{(p-k)!(p+k+1)!} =
\sum_{k=0}^p \frac{(-1)^k(1-\lambda)^{2k+1}(1+\lambda)^{2p-2k}}{
k!(p-k)!(2k+1)} = \frac{1}{p!}\int_0^1 ( (1+\lambda)^2-t^2(1-\lambda)^2 )^p dt
\ee
At $\lambda = 0$, this identity becomes just a B-function evaluation

\be
\dfrac{4^p p!}{(2p+1)!} = \sum_{k=0}^p \frac{(-1)^k}{
k!(p-k)!(2k+1)} = \frac{1}{p!} \int_0^1 ( 1 - t^2 )^p dt
\ee
while for generic $\lambda \neq 0$ it can be considered as a hypergeometric identity and proven, say, by the systematic methods of Wilf and Zeilberger \cite{A=B}.

The even part of the contour integral can be treated similarly. As one can see, a quite complicated contour integral gives rise to a seemingly no-less-complicated series expansion, which -- after the Harer-Zagier transform! -- turns into an elementary function. However, we still do not posess any general expression for these elementary functions; even derivation of the simplest examples, as it is clear from the above calculation, is not quite easy. For higher-point functions, such calculations become increasingly more complicated, and remain a challenge.

\section*{Appendix: Okounkov's formula for cubic Kontsevich model \cite{Ok}}

Another important direction of generalization is from the Gaussian Kontsevich model to Generalized Kontsevich models of higher degree. In the case of cubic Kontsevich model

\be
Z_K= {\int DX\ \exp\left(-{1\over
3}\hbox{Tr}X^3-{1\over\sqrt{3}}\hbox{Tr}AX^2\right)\over \int DX\
\exp\left(-{1\over\sqrt{3}}\hbox{Tr}AX^2\right)}, \ \ \ \ \tau_{2k+1}\equiv \displaystyle{}{3^{2k+1}\over 2k+1}\hbox{Tr}
A^{-2k-1} 
\ee
where the conventional resolvents have a form

\be
\rho_K(z_1,\ldots,z_m)={\hat \Delta}(z_1) \ldots {\hat \Delta}(z_m) \log Z_K \Big|_{\tau = 0}, \ \ \ \ {\hat \Delta}(z)= \sum_{n = 
0}^{\infty}\frac{1}{z^{n+3/2}}\frac{\partial}{\partial \tau_{2n+1}} 
\ee 
a contour integral representation was obtained by Okounkov (see \cite{Ok} for more details) for the exponential functions $e_K(x_1,\ldots,x_k)$, which are connected with the above resolvents by Laplace transform:

\begin{equation}
 \rho_K(z_1,\ldots,z_k)=2^k\, \int\limits_0^\infty \prod_{i=1}^k ds_i e^{-\sum\limits_i s_iz_i} e_K(s_1,\ldots,s_k) \label{rho_eta}
\end{equation}
This contour integral representation has a form 

\be
e_K(s) = {\cal E}(s), \nn \\
e_K(s_1,s_2) = {\cal E}(s_1,s_2) - {\cal E}(s_1){\cal E}(s_2), \nn \\
e_K(s_1,s_2,s_3) = {\cal E}(s_1,s_2,s_3) -
{\cal E}(s_1){\cal E}(s_2,s_3) - {\cal E}(s_2){\cal E}(s_1,s_3)
- {\cal E}(s_3){\cal E}(s_1,2)
+ {\cal E}(s_1){\cal E}(s_2){\cal E}(s_3), \nn \\
\ldots
\ee
and so on, where
\be
{\cal E}(s_1,\ldots,s_n) = \frac{1}{2^n\pi^{n/2}}
\prod_{\alpha=1}^n \frac{e^{\frac{1}{12}s_\alpha^3}}{\sqrt{s_\alpha}}
\int_{t_\alpha \geq 0} dt_\alpha \exp\left(
-\frac{(t_\alpha-t_{\alpha+1})^2}{4s_\alpha}
- \frac{(t_\alpha+t_{\alpha+1})s_\alpha}{2}\right)
\ee
These formulas can be considered as direct generalisations of Brezin-Hikami contour integrals for the Gaussian Kontsevich model. It is puzzling to ask, whether similar integral representations exist for arbitrary monomial potentials in GKM, or maybe even for generic non-monomial potentials. Of interest is also to find analogues of elementary multidensities in these models.

\section*{Acknowledgements}

Our work is partly supported by Russian Federal Nuclear Energy Agency, Federal Agency for Science and
Innovations of Russian Federation under contract 02.740.11.5194, by RFBR grants 10-01-00536, by joint grants 09-02-90493-Ukr, 09-02-93105-CNRSL, 09-01-92440-CE, 09-02-91005-ANF, 10-02-92109-Yaf-a, by CNRS (A.M.) and by Dynasty Foundation (Sh.Sh.).


\begin{thebibliography}{12}

\bibitem{UFN3}
A.Morozov,
Phys.Usp.(UFN) {\bf 35} (1992) 671-714; {\bf 37} (1994) 1, hep-th/9303139;
hep-th/9502091; hep-th/0502010 \\
A.Mironov, Int.J.Mod.Phys. {\bf A9} (1994) 4355, hep-th/9312212; Phys.Part.Nucl.
{\bf 33} (2002) 537; hep-th/9409190; Theor.Math.Phys. {\bf 114} (1998) 127, q-alg/9711006
\bibitem{AMM1}
A.Alexandrov, A.Mironov and A.Morozov, Int.J.Mod.Phys. {\bf A19} (2004) 4127, hep-th/0310113
\bibitem{HZ}
J. Harer, D. Zagier, Inv. Math. \textbf{85} (1986) 457-485
\bibitem{MS_HZ}
Sh.Shakirov and A.Morozov, JHEP 0912(2009)003, arXiv:0906.0036

\bibitem{GKM}
M.~Kontsevich, Commun.\ Math.\ Phys.\  {\bf 147} (1992) 1; Funk.\ Anal.\ i Pril.\  {\bf 25} (1991) 50;
\\
  S.~Kharchev, A.~Marshakov, A.~Mironov, A.~Morozov and A.~Zabrodin,
  Phys.\ Lett.\  B {\bf 275} (1992) 311,
 hep-th/9111037;  Nucl.\ Phys.\  B {\bf 380} (1992) 181,
  hep-th/9201013;
\\
  C.~Itzykson and J.~B.~Zuber,
  Int.\ J.\ Mod.\ Phys.\  A {\bf 7} (1992) 5661,
  hep-th/9201001

\bibitem{dua}
L.Chekhov and Yu.Makeenko, Phys.Lett. B278 (1992) 271-278, arXiv:hep-th/9202006 \\
  S.~Kharchev and A.~Marshakov, hep-th/9210072;
\\
  S.~Kharchev and A.~Marshakov, Int.\ J.\ Mod.\ Phys.\  A {\bf 10} (1995) 1219, hep-th/9303100
  \\
A.~Mironov, A.~Morozov and G.~W.~Semenoff,
  Int.\ J.\ Mod.\ Phys.\  A {\bf 11} (1996) 5031, hep-th/9404005
  \\
A.Alexandrov, A.Mironov and A.Morozov, JHEP 0912(2009)053, arXiv:0906.3305

\bibitem{BH}
E.Brezin and S.Hikami, JHEP 0710(2007)096, arXiv:0709.3378; Commun.Math.Phys.283(2008)507-521, arXiv:0708.2210; arXiv:cond-mat/9804024

\bibitem{LMNS}
N.Nekrasov, Adv.Theor.Math.Phys. {\bf 7} (2004) 831-864 \\
G.Moore, N.Nekrasov, S.Shatashvili, Nucl.Phys. {\bf B534} (1998) 549-611, hep-th/9711108;
hep-th/9801061 \\
A.Losev, N.Nekrasov and S.Shatashvili, Commun.Math.Phys. {\bf 209} (2000) 97-121, hep-th/9712241;
ibid. 77-95, hep-th/9803265

\bibitem{AGT}
D.Gaiotto, arXiv:0908.0307 \\
 L.Alday, D.Gaiotto and Y.Tachikawa,
Lett.Math.Phys. {\bf 91} (2010) 167-197, arXiv:0906.3219 \\
N.Wyllard, JHEP \textbf{0911} (2009) 002, arXiv:0907.2189 \\
D.Nanopoulos and D.Xie, arXiv:0908.4409; JHEP \textbf{1003}(2010)043, arXiv:0911.1990  \\
A.Marshakov, A.Mironov and A. Morozov, arXiv:0907.3946; Phys.Lett.\textbf{B682} (2009) 125-129, arXiv:0909.2052; JHEP \textbf{0911} (2009) 048, arXiv:0909.3338 \\
A. Mironov and A. Morozov, Nucl.Phys.\textbf{B825}(2010)1-37, arXiv:0908.2569; Phys.Lett.\textbf{B682}(2009)118-124, arXiv:0909.3531  \\
N.Nekrasov and S.Shatashvili, Nucl.Phys.B, Proc.Suppl. \textbf{192-193} (2009) 91-112; arXiv:0901.4748, arXiv:0908.4052  \\
N.Nekrasov and E.Witten

\bibitem{AGTdua}
A.Mironov, A.Morozov and And.Morozov, arXiv:1003.5752

\bibitem{DFMM}
Vl.Dotsenko and V.Fateev, Nucl.Phys. {\bf B240} (1984) 312-348 \\
A.Gerasimov, A.Marshakov, A.Morozov, M.Olshanetsky,
S. Shatashvili,
%\emph{Wess-Zumino-Witten model as a theory of free fields},
Int.J.Mod.Phys. {\bf A5} (1990) 2495 \\
A.Gerasimov, A.Marshakov and A.Morozov,
%\emph{ Free Field Representation Of Parafermions And Related Coset Models},
Nucl.Phys. {\bf B328} (1989) 664, Theor.Math.Phys. {\bf 83} (1990) 466-473;
%\emph{Hamiltonian Reduction Of Wess-Zumino-Witten Theory
%From The Point Of View Of Bosonization},
Phys.Lett. {\bf B236} (1990) 269, Sov.J.Nucl.Phys. {\bf 51} (1990) 371-372 \\
A. Marshakov, A. Mironov, and A. Morozov, Phys.Lett.B265(1991)99-107 \\
S. Kharchev, A. Marshakov, A. Mironov, A. Morozov and S. Pakuliak, Nucl.Phys.B404 (1993) 717-750,  hep-th/9208044 \\
A.Mironov and A. Morozov, Phys.Lett.\textbf{B680}(2009)188-194, arXiv:0908.2190 \\
R. Dijkgraaf and C. Vafa, arXiv:0909.2453 \\
H.Itoyama, K.Maruyoshi and T.Oota, arXiv:0911.4244 \\
T. Eguchi and K. Maruyoshi, arXiv:0911.4797 \\
R. Schiappa and N. Wyllard, arXiv:0911.5337 \\
A.Mironov, A.Morozov and Sh.Shakirov, JHEP \textbf{1002} (2010) 030, arXiv:0911.5721; arXiv:1001.0563; arXiv:1004.2917

\bibitem{Ok}
A.Okounkov, arXiv:math/0101201 \\
A.Alexandrov, A.Mironov, A.Morozov and P.Putrov, Int.J.Mod.Phys.\textbf{A24}(2009)4939-4998, arXiv:0811.2825

\bibitem{Nek}
N.Nekrasov, Adv.Theor.Math.Phys.\textbf{7} (2004) 831-864, hep-th/0206161 \\
R.Flume and R.Poghossian, Int.J.Mod.Phys. \textbf{A18} (2003) 2541, hep-th/0208176 \\
S. Shadchin, arXiv:hep-th/0601167

\bibitem{A=B}M. Petkovsek, H. Wilf and D. Zeilberger, \emph{"A = B"}, 1996.

\end{thebibliography}
\end{document}